\documentclass{llncs}

\usepackage{t1enc}
\usepackage{amsmath}
\usepackage{latexsym}
\usepackage{theorem}
\usepackage{amssymb}
\usepackage{url}

\newcommand{\comment}[1]{}
\newcommand{\bu}{$\bullet$}

\newcommand{\vsp}[1][3mm]{\vspace*{#1}}
\newcommand{\moins}{\setminus}
\newcommand{\vide}{\emptyset}
\newcommand{\ie}{{\em i.e.} }

\newcommand{\pos}{\mr{Pos}}

\renewcommand{\a}{\rightarrow}

\newcommand{\ad}{\downarrow}

\renewcommand{\to}{\mapsto}
\newcommand{\al}{\leftarrow}

\newcommand{\et}{\wedge}

\newcommand{\sle}{\subseteq}

\newcommand{\h}[1]{{\widehat{#1}}}

\newcommand{\g}{\gamma}
\newcommand{\G}{\Gamma}

\renewcommand{\t}{\theta}

\newcommand{\la}{\lambda}

\renewcommand{\r}{\rho}
\newcommand{\s}{\sigma}
\renewcommand{\S}{\Sigma}

\newcommand{\mi}{\mathit}
\newcommand{\mc}{\mathcal}

\newcommand{\mr}{\mathrm}

\newcommand{\cA}{\mc{A}}

\newcommand{\cC}{\mc{C}}

\newcommand{\cE}{\mc{E}}
\newcommand{\cF}{\mc{F}}

\newcommand{\cR}{\mc{R}}
\newcommand{\cS}{\mc{S}}
\newcommand{\cT}{\mc{T}}

\newcommand{\cX}{\mc{X}}

\newenvironment{rul}
  {$\begin{array}{rcl}}
  {\end{array}$}

\newenvironment{rew}[1][~~\a~~]
  {$\begin{array}{r@{#1}l}}
  {\end{array}$}
\newenvironment{rewc}[1][~~\a~~]
  {\begin{center}\begin{rew}[#1]}
  {\end{rew}\end{center}}

  {\end{array}$\end{center}}

\newcounter{counter}

\newcounter{explnum}
{\theorembodyfont{\rmfamily} 
  \newtheorem{dfn}[counter]{Definition}
  
  \newtheorem{thm}[counter]{Theorem}

  \newtheorem{expl}[explnum]{Example}
}

\newcommand{\cqfd}{\hfill$\blacksquare$} 
\newenvironment{prf}{{\bf Proof.}}{}

\newenvironment{lstgeneric}[2]
  {\begin{list}{#1}{\topsep=.5mm\itemsep=.5mm\parsep=0mm%
    \itemindent=-3ex\labelsep=1ex\labelwidth=0ex #2}}
  {\end{list}}

\newenvironment{lst}[1]
  {\begin{lstgeneric}{#1}{\itemindent=-1ex}}
  {\end{lstgeneric}}

\newenvironment{enumi}[1]
  {\begin{lstgeneric}{}{\usecounter{enumi}\leftmargin=7mm%
    }}
  {\end{lstgeneric}}

\newcommand{\comb}{\mi{comb}}
\newcommand{\sort}{\mi{sort}}
\newcommand{\norm}{\mi{norm}}
\newcommand{\orient}{\mi{orient}}
\newcommand{\leaves}{\mi{leaves}}
\newcommand{\step}{\mi{step}}

\newcommand{\ocaml}{OCaml}
\newcommand{\moca}{Moca}
\newcommand{\focal}{Focal}
\newcommand{\coq}{Coq}
\newcommand{\cime}{CiME}
\newcommand{\zenon}{Zenon}

\newcommand{\vpos}{\mr{VPos}}

\begin{document}


\title{On the implementation of construction functions for non-free concrete data types}

\author{Fr\'ed\'eric Blanqui\inst{1}
\and Th\'er\`ese Hardin\inst{2}
\and Pierre Weis\inst{3}}

\institute{INRIA \& LORIA, BP 239,
54506 Villers-l\`es-Nancy Cedex, France
\and UPMC,
LIP6, 104, Av. du Pr. Kennedy, 75016 Paris, France
\and INRIA, Domaine de Voluceau, BP 105,
78153 Le Chesnay Cedex, France}

\maketitle

\begin{abstract}
Many algorithms use concrete data types with some additional invariants. The
set of values satisfying the invariants is often a set of representatives for
the equivalence classes of some equational theory. For instance, a sorted list
is a particular representative wrt commutativity. Theories like associativity,
neutral element, idempotence, etc. are also very common. Now, when one wants to
combine various invariants, it may be difficult to find the suitable
representatives and to efficiently implement the invariants. The preservation
of invariants throughout the whole program is even more difficult and error
prone. Classically, the programmer solves this problem using a combination of
two techniques: the definition of appropriate construction functions for the
representatives and the consistent usage of these functions ensured via
compiler verifications. The common way of ensuring consistency is to use an
abstract data type for the representatives; unfortunately, pattern matching on
representatives is lost. A more appealing alternative is to define a concrete
data type with private constructors so that both compiler verification and
pattern matching on representatives are granted. In this paper, we detail the
notion of private data type and study the existence of construction
functions. We also describe a prototype, called \moca, that addresses the
entire problem of defining concrete data types with invariants: it generates
efficient construction functions for the combination of common invariants and
builds representatives that belong to a concrete data type with private
constructors.
\end{abstract}

\section{Introduction}

Many algorithms use data types with some additional invariants. Every
function creating a new value from old ones must be defined so that
the newly created value satisfy the invariants whenever the old ones
so do.

One way to easily maintain invariants is to use abstract data types
(ADT): the implementation of an ADT is hidden and construction and
observation functions are provided. A value of an ADT can only be
obtained by recursively using the construction functions. Hence, an
invariant can be ensured by using appropriate construction
functions. Unfortunately, abstract data types preclude pattern
matching, a very useful feature of modern programming languages
\cite{ocaml3.09,haskell03,moreau03cc,tom2.3}. There have been various
attempts to combine both features in some way.

In \cite{wadler87popl}, P. Wadler proposed the mechanisms of {\em
views}. A view on an ADT $\alpha$ is given by providing a concrete
data type (CDT) $\gamma$ and two functions $in:\alpha\a\g$ and
$out:\g\a\alpha$ such that $in\circ out=id_\g$ and $out\circ
in=id_\alpha$. Then, a function on $\alpha$ can be defined by matching
on $\g$ (by implicitly using $in$) and the values of type $\g$
obtained by matching can be injected back into $\alpha$ (by implicitly
using $out$). However, by leaving the applications of $in$ and $out$
implicit, we can easily get inconsistencies whenever $in$ and $out$
are not inverses of each other. Since it may be difficult to satisfy
this condition (consider for instance the translations between
cartesian and polar coordinates), these views have never been
implemented. Following the suggestion of W. Burton and R. Cameron to
use the $in$ function only \cite{burton93jfp}, some propositions have
been made for various programming languages but none has been
implemented yet \cite{burton96,okasaki98ml}.

In \cite{burton93jfp}, W. Burton and R. Cameron proposed another very
interesting idea which seems to have attracted very little
attention. An ADT must provide construction and observation
functions. When an ADT is implemented by a CDT, they propose to also
export the constructors of the CDT but only for using them as patterns
in pattern matching clauses. Hence, the constructors of the underlying
CDT can be used for pattern matching but not for building values: only
the construction functions can be used for that purpose. Therefore,
one can both ensure some invariants and offer pattern matching. These
types have been introduced in \ocaml\ by the third author
\cite{weis03} under the name of {\em concrete data type with private
constructors}, or {\em private data type} (PDT) for short.\vsp[2mm]

Now, many invariants on concrete data types can be related to some
equational theory. Take for instance the type of $list$ with the
constructors $[]$ and $::$. Given some elements $v_1..v_n$, the sorted
list which elements are $v_1..v_n$ is a particular representative of
the equivalence class of $v_1$::..::$v_n$::$[]$ modulo the equation
$x$::$y$::$l$=$y$::$x$::$l$. Requiring that, in addition, the list
does not contain the same element twice is a particular representative
modulo the equation $x$::$x$::$l$=$x$::$l$.

Consider now the type of join lists with the constructors $empty$,
$singleton$ and $append$, for which concatenation is of constant
complexity. Sorting corresponds to associativity and commutativity of
$append$. Requiring that no argument of $append$ is $empty$
corresponds to neutrality of $empty$ wrt $append$. We have a structure
of commutative monoid.

More generally, given some equational theory on a concrete data type,
one may wonder whether there exists a representative for each
equivalence class and, if so, whether a representative of $C(t_1\ldots
t_n)$ can be efficiently computed knowing that $t_1\ldots t_n$ are
themselves representatives.

In \cite{thompson86lfp,thompson90scp}, S. Thompson describes a
mechanism introduced in the Miranda functional programming language
for implementing such non-free concrete data types without precluding
pattern matching. The idea is to provide conditional rewrite rules,
called {\em laws}, that are implicitly applied as long as possible on
every newly created value. This can also be achieved by using a PDT
which construction functions (primed constructors in
\cite{thompson86lfp}) apply as long as possible each of the laws.
Then, S. Thompson studies how to prove the correctness of functions
defined by pattern matching on such {\em lawful types}. However, few
hints are given on how to check whether the laws indeed implement the
invariants one has in mind. For this reason and because reasoning on
lawful types is difficult, the law mechanism was removed from
Miranda.\vsp[2mm]

In this paper, we propose to specify the invariants by unoriented
equations (instead of rules). We will call such a type a {\em
relational data type} (RDT). Sections \ref{sec-pdt} and \ref{sec-rdt}
introduce private and relational data types. Then, we study when an
RDT can be implemented by a PDT, that is, when there exist
construction functions computing some representative for each
equivalence class. Section \ref{sec-ex} provides some general
existence theorem based on rewriting theory. But rewriting may be
inefficient. Section \ref{sec-genr} provides, for some common
equational theories, construction functions more efficient than the
ones based on rewriting. Section \ref{sec-moca} presents \moca, an
extension of \ocaml\ with relational data types whose construction
functions are automatically generated. Finally, Section
\ref{sec-futur} discusses some possible extensions.


\section{Concrete data types with private constructors}
\label{sec-pdt}


We first recall the definition of a first-order term algebra. It will
be useful for defining the values of concrete and private data types.

\begin{dfn}[First-order term algebra]
A {\em sorted term algebra definition} is a triplet $\cA=(\cS,\cC,\S)$
where $\cS$ is a non-empty set of {\em sorts}, $\cC$ is a non-empty
set of {\em constructor symbols} and $\S:\cC\a\cS^+$ is a {\em
signature} mapping a non-empty sequence of sorts to every constructor
symbol. We write $C:\s_1 \ldots\s_n\s_{n+1}\in\S$ to denote the fact
that $\S(C)=\s_1 \ldots\s_n\s_{n+1}$. Let $\cX=(\cX_\s)_{\s\in\cS}$ be
a family of pairwise disjoint sets of {\em variables}. The sets
$\cT_\s(\cA,\cX)$ of {\em terms of sort $\s$} are inductively defined
as follows:

\begin{lst}{--}
\item If $x\in\cX_\s$, then $x\in\cT_\s(\cA,\cX)$.
\item If $C:\s_1\ldots\s_{n+1}\in\S$ and
$t_i\in\cT_{\s_i}(\cA,\cX)$, then
${C(t_1,\ldots,t_n)}\in\cT_{\s_{n+1}}(\cA,\cX)$.
\end{lst}

\noindent
Let $\cT_\s(\cA)$ be the set of terms of sort $\s$ containing no
variable.
\end{dfn}

In the following, we assume given a set $\cS_0$ of primitive types
like {\tt int}, {\tt string}, \ldots and a set $\cC_0$ of primitive
constants {\tt 0}, {\tt 1}, {\tt "foo"}, \ldots  Let $\S_0$ be the
corresponding signature ($\S_0({\tt 0})={\tt int}$, \ldots).


In this paper, we call {\em concrete data type} (CDT) an inductive
type {\em\`a la} ML defined by a set of {\em constructors}. More
formally:

\begin{dfn}[Concrete data type]
A {\em concrete data type definition} is a triplet $\G=(\g,\cC,\S)$
where $\g$ is a sort, $\cC$ is a non-empty set of {\em constructor
symbols} and $\S:\cC\a(\cS_0\cup\{\g\})^+$ is a {\em signature} such
that, for all $C\in\cC$, $\S(C)=\s_1..\s_n\g$. The set $Val(\g)$ of
{\em values of type $\g$} is the set of terms $\cT_\g(\cA_\G)$ where
$\cA_\G=(\cS_0\cup\{\g\},\cC_0\cup\cC,\S_0\cup\S)$.
\end{dfn}

This definition of CDTs corresponds to a small but very useful subset
of all the possible types definable in ML-like programming
languages. For the purpose of this paper, it is not necessary to use a
more complex definition.

\begin{expl}
The following type\footnote{Examples are written with
\ocaml\ \cite{ocaml3.09}, they can be readily translated in any
programming language offering pattern-matching with textual priority,
as Haskell, SML, etc.} {\tt cexp} is a CDT definition with two
constant constructors of sort {\tt cexp} and a binary operator of sort
{\tt cexp} {\tt cexp} {\tt cexp}.

{\small\begin{verbatim}
type cexp =  Zero | One | Opp of cexp | Plus of cexp * cexp
\end{verbatim}}
\end{expl}


Now, a private data type definition is like a CDT definition together
with construction functions as in abstract data types. Constructors
can be used as patterns as in concrete data types but they {\em
cannot} be used for value creation (except in the definition of
construction functions). For building values, one must use
construction functions as in abstract data types. Formally:

\begin{dfn}[Private data type]
A {\em private data type definition} is a pair $\Pi=(\G,\cF)$ where
$\G=(\pi,\cC,\S)$ is a CDT definition and $\cF$ is a family of {\em
construction functions} $(f_C)_{C\in\cC}$ such that, for all
$C:\s_1..\s_n\pi\in\S$,
$f_C:\cT_{\s_1}(\cA_\G)\times\ldots\times\cT_{\s_n}(\cA_\G)\a\cT_\pi(\cA_\G)$.
Let $Val(\pi)$ be the set of the {\em values of type $\pi$}, that is,
the set of terms that one can build by using the construction
functions only. The function $f:\cT_\pi(\cA_\G)\a\cT_\pi(\cA_\G)$ such
that, for all $C:\s_1..\s_n\pi\in\S$ and $t_i\in\cT_{\s_i}(\cA_\G)$,
$f(C(t_1..t_n))=f_C(f(t_1)..f(t_n))$, is called the {\em normalization
function associated to $\cF$}.
\end{dfn}


This is quite immediate to see that:

\begin{lemma}
$Val(\pi)$ is the image of $f$.
\end{lemma}


PDTs have been implemented in \ocaml\ by the third author
\cite{weis03}.  Extending a programming language with PDTs is not very
difficult: one only needs to modify the compiler to parse the PDT
definitions and check that the conditions on the use of constructors
are fulfilled.

Note that construction functions have no constraint in general: the
full power of the underlying programming language is available to
define them.


It should also be noted that, because the set of values of type $\pi$
is a subset of the set of values of the underlying CDT $\g$, a
function on $\pi$ defined by pattern matching may be a total function
even though it is not defined on all the possible cases of
$\g$. Defining a function with patterns that match no value of type
$\pi$ does not harm since the corresponding code will never be run. It
however reveals that the developer is not aware of the distinction
between the values of the PDT and those of the underlying CDT, and
thus can be considered as a programming error. To avoid this kind of
errors, it is important that a PDT comes with a clear identification
of its set of possible values. To go one step further, one could
provide a tool for checking the completeness and usefulness of
patterns that takes into account the invariants, when it is
possible. We leave this for future work.


\begin{expl}
Let us now start our running example with the type {\tt exp}
describing operations on arithmetic expressions.

{\small\begin{verbatim}
type exp = private Zero | One | Opp of exp | Plus of exp * exp
\end{verbatim}}

This type {\tt exp} is indeed a PDT built upon the CDT {\tt
cexp}. Prompted by the keyword {\tt private}, the \ocaml\ compiler
forbids the use of {\tt exp} constructors (outside the module {\tt
my\_exp.ml} containing the definition of {\tt exp}) except in
patterns.  If {\tt Zero} is supposed to be neutral by the writer of
{\tt my\_exp.ml}, then he/she will provide construction functions as
follows:

{\small\begin{verbatim}
let rec zero = Zero and one = One and opp x = Opp x
and plus = function
| (Zero,y) ->  y
| (y,Zero) -> y
| (x,y) -> Plus(x,y)
\end{verbatim}}
\end{expl}

\comment{Let us now give a very simple example showing how PDTs can be used for
representing subsets:\footnote{For our examples, we use the \ocaml\
\cite{ocaml3.09} syntax but they could easily be translated to other
programming languages.}

{\small\begin{verbatim}
(* interface file nat.mli *)
type nat = private Nat of int
val nat_of_int : int -> nat

(* implementation file nat.ml *)
type nat = Nat of int
let nat_of_int x = if x<0 then invalid_arg "negative" else Nat x

(* interface file array.mli *)
open Nat
type array
val array_make : nat -> string -> array

(* implementation file array.ml *)
open Nat
type array = ...
let int_of_nat (Nat x) = x
let array_make n s = ... (* here, we have [int_of_nat n >= 0] *)
\end{verbatim}}

A value $n$ of type {\tt nat} can be obtained only by using {\tt
int\_of\_nat} which ensures that $n$ is a non-negative integer. Note
that the test $n\ge 0$ is done only once, at value creation.
We think that it should be possible to get rid of the constructor {\tt
Nat} and allow declarations like {\tt type nat = private int}.
}

\section{Relational data types}
\label{sec-rdt}

We mentioned in the introduction that, often, the invariants upon
concrete data types are such that the set of values satisfying them is
indeed a set of representatives for the equivalence classes of some
equational theory. We therefore propose to specify invariants by a set
of unoriented equations and study to which extent such a specification
can be realized with an abstract or private data type. In case of a
private data type however, it is important to be able to describe the
set of possible values.

\begin{dfn}[Relational data type]
A {\em relational data type (RDT) definition} is a pair $(\G,\cE)$
where $\G=(\pi,\cC,\S)$ is a CDT definition and $\cE$ is a finite set
of equations on $\cT_\pi(\cA_\G,\cX)$. Let $=_\cE$ be the smallest
congruence relation containing $\cE$. Such an RDT is {\em
implementable} by a PDT $(\G,\cF)$ if the family of construction
functions $\cF=(f_C)_{C\in\cC}$ is {\em valid wrt $\cE$}:

\begin{lst}{--}
\item [\bf(Correctness)]
For all $C:\s_1..\s_n\pi$ and $v_i\in Val(\s_i)$,
$f_C(v_1..v_n)=_\cE C(v_1..v_n)$.
\item [\bf(Completeness)]
For all $C:\s_1..\s_n\s$, $v_i\in Val(\s_i)$,
$D:\tau_1..\tau_p\s\in\S$ and $w_i\in Val(\tau_i)$, $f_C(v_1..v_n)=
f_D(w_1..w_p)$ whenever $C(v_1..v_n)=_\cE D(w_1..w_p)$.
\end{lst}
\end{dfn}


We are going to see that the existence of a valid family of
construction functions is equivalent to the existence of a valid
normalization function:

\begin{dfn}[Valid normalization function]
A map $f:\cT_\pi(\cA_\G)\a\cT_\pi(\cA_\G)$ is a {\em valid
normalization function} for an RDT $(\G,\cE)$ with $\G=(\pi,\cC,\S)$
if:

\begin{lst}{--}
\item [\bf(Correctness)]
For all $t\in\cT_\pi(\cA_\G)$, $f(t)=_\cE t$.
\item [\bf(Completeness)]
For all $t,u\in\cT_\pi(\cA_\G)$, $f(t)= f(u)$ whenever $t=_\cE u$.
\end{lst}
\end{dfn}

Note that a valid normalization function is idempotent ($f\circ f=f$)
and provides a decision procedure for $=_\cE$ (the boolean function
$\la xy.f(x)= f(y)$).


\begin{thm}
\label{thm-f}
The normalization function associated to a valid family is a valid
normalization function.
\end{thm}

\begin{prf}
\begin{lst}{--}
\item Correctness. We proceed by induction on the size of $t\in\cT_\pi$.
We have $C:\s_1..\s_n\pi\in\S$ and $t_i$ such that $t=C(t_1..t_n)$. By
definition, $f(t)=f_C(f(t_1)..$ $f(t_n))$. By induction hypothesis,
$f(t_i)=_\cE t_i$. Since the family is valid and $f(t_1)..f(t_n)$ are
values, $f_C(f(t_1)..f(t_n))=_\cE C(f(t_1)..f(t_n))$. Thus, $f(t)=_\cE
t$.

\item Completeness.
Let $t,u\in\cT_\pi$ such that $t=_\cE u$. We have $t=C(t_1..t_n)$ and
$u=D(u_1..u_p)$. By definition, $f(t)=f_C(f(t_1)..f(t_n))$ and
$f(u)=f_D(f(u_1)..f(u_p))$. By correctness, $f(t_i)=_\cE t_i$ and
$f(u_j)=_\cE u_j$. Hence, $C(f(t_1)..f(t_n))=_\cE
D(f(u_1)..f(u_p))$. Since the family is valid and $f(t_1)..f(t_n)$ are
values, $f_C(f(t_1)$ $..f(t_n))=f_D(f(t_1)..f(t_n))$. Thus,
$f(t)=f(u)$.\cqfd\\
\end{lst}
\end{prf}


Conversely, given $f:\cT_\pi(\cA_\G)\a\cT_\pi(\cA_\G)$, one can
easily define a family of construction functions that is valid
whenever $f$ is a valid normalization function.

\begin{dfn}[Associated family of constr. functions]
Given a CDT $\G=(\pi,\cC,\S)$ and a function
$f:\cT_\pi(\cA_\G)\a\cT_\pi(\cA_\G)$, the {\em family of construction
functions associated to $f$} is the family $(f_C)_{C\in\cC}$ such
that, for all $C:\s_1..\s_n\pi\in\S$ and $t_i\in\cT_{\s_1}(\cA_\G)$,
$f_C(t_1, \ldots,t_n)=f(C(t_1, \ldots, t_n))$.
\end{dfn}

\begin{thm}
\label{thm-g}
The family of construction functions associated to a valid
normalization function is valid.
\end{thm}

\comment{
\begin{prf}
\begin{lst}{--}
\item Correctness.
Let $C:\s_1..\s_n\g\in\S$ and $v_i\in Val_\cF(\s_i)$. Since $f$ is
valid, we have $f_C(v_1, \ldots, v_n)=f(C(v_1, \ldots, v_n))=_\cE
C(v_1, \ldots, v_n)$.

\item Completeness.
Let $C:\s_1..\s_n\pi\in\S$, $v_i\in Val_\cF(\s_i)$,
$D:\tau_1..\tau_p\pi\in\S$ and $w_i\in Val_\cF(\s_i)$ such that
$C(v_1, \ldots, v_n)=_\cE D(w_1, \ldots, w_p)$. Since $f$ is valid,
$f_C(v_1, \ldots, v_n)=f(C(v_1, \ldots, v_n))= f(D(w_1, \ldots,
w_p))=f_D(w_1, \ldots, w_p)$.\cqfd
\end{lst}
\end{prf}
}

\begin{expl}
We can choose {\tt cexp} as the underlying CDT and $\cE = \{$ {\tt
Plus x Zero = x}$\}$ to define a RDT implementable by the PDT {\tt
exp}, with the valid family of construction functions {\tt zero}, {\tt
one}, {\tt opp}, {\tt plus}.
\end{expl}


\section{On the existence of construction functions}
\label{sec-ex}

In this section, we provide a general theorem for the existence of
valid families of construction functions based on rewriting theory. We
recall the notions of rewriting and completion. The interested reader
may find more details in \cite{dershowitz90book}.\vsp[1mm]

\noindent
{\bf Standard rewriting.} A {\em rewrite rule} is an ordered pair of
terms $(l,r)$ written $l\a r$. A rule is {\em left-linear} if no
variable occurs twice in its left hand side $l$.

As usual, the set $\pos(t)$ of {\em positions in $t$} is defined as a
set of words on positive integers. Given $p\in\pos(t)$, let $t|_p$ be
the subterm of $t$ at position $p$ and $t[u]_p$ be the term $t$ with
$t|_p$ replaced by $u$.

Given a finite set $\cR$ of rewrite rules, the {\em rewriting
relation} is defined as follows: $t\a_\cR u$ iff there are
$p\in\pos(t)$, $l\a r\in\cR$ and a substitution $\t$ such that
$t|_p=l\t$ and $u=t[r\t]_p$. A term $t$ is an {\em $\cR$-normal form}
if there is no $u$ such that $t\a_\cR u$. Let $=_\cR$ be the
symmetric, reflexive and transitive closure of $\a_\cR$.

A {\em reduction ordering} $\succ$ is a well-founded ordering (there
is no infinitely decreasing sequence $t_0\succ t_1\succ\ldots$) stable
by context ($C(..t..)\succ C(..u..)$ whenever $t\succ u$) and
substitution ($t\t\succ u\t$ whenever $t\succ u$). If $\cR$ is
included in a reduction ordering, then $\a_\cR$ is well-founded
(terminating, strongly normalizing).

We say that $\a_\cR$ is {\em confluent} if, for all terms $t,u,v$ such
that $u\al^*_\cR t\a^*_\cR v$, there exists a term $w$ such that
$u\a^*_\cR w\al^*_\cR v$. This means that the relation
$\al^*_\cR\a^*_\cR$ is included in the relation $\a^*_\cR\al^*_\cR$
(composition of relations is written by juxtaposition).

If $\a_\cR$ is confluent, then every term has at most one normal
form. If $\a_\cR$ is well-founded, then every term has at least one
normal form. Therefore, if $\a_\cR$ is confluent and terminating, then
every term has a unique normal form.\vsp[1mm]


\noindent
{\bf Standard completion.} Given a finite set $\cE$ of equations and a
reduction ordering $\succ$, the standard Knuth-Bendix completion
procedure \cite{bendix70book} tries to find a finite set $\cR$ of
rewrite rules such that:

\begin{lst}{\bu}
\item $\cR$ is included in $\succ$,
\item $\a_\cR$ is confluent,
\item $\cR$ and $\cE$ have same theory: ${=_\cE}={=_\cR}$.
\end{lst}

Note that completion may fail or not terminate but, in case of
successful termination, $\cR$-normalization provides a decision
procedure for $=_\cE$ since $t=_\cE u$ iff the $\cR$-normal forms of
$t$ and $u$ are syntactically equal.\vsp[2mm]


However, since permutation theories like commutativity or
associativity and commutativity together (written AC for short) are
included in no reduction ordering, dealing with them requires to
consider rewriting with pattern matching modulo these theories and
completion modulo these theories. In this paper, we restrict our
attention to AC.

\begin{dfn}[Associative-commutative equations]
Let $Com$ be the set of commutative constructors, \ie the set of
constructors $C$ such that $\cE$ contains an equation of the form
$C(x,y)=C(y,x)$. Then, let $\cE_{AC}$ be the subset of $\cE$ made of
the commutativity and associativity equations for the commutative
constructors, $=_{AC}$ be the smallest congruence relation containing
$\cE_{AC}$ and $\cE_{\neg AC}=\cE\moins\cE_{AC}$.
\comment{
\begin{center}
$\begin{array}{rl}
\cE_{AC}= & \{C(x,y)=C(y,x)\in\cE~|~C\in Com\}\\
& \cup~\{C(x,C(y,z))=C(C(x,y),z)\in\cE~|~C\in Com\}\\
\end{array}$
\end{center}
}
\end{dfn}

\noindent
{\bf Rewriting modulo AC.} Given a set $\cR$ of rewrite rules, {\em
rewriting with pattern matching modulo $AC$} is defined as follows:
$t\a_{\cR,AC} u$ iff there are $p\in\pos(t)$, $l\a r\in\cR$ and a
substitution $\t$ such that $t|_p=_{AC}l\t$ and $u=t[r\t]_p$. A
reduction ordering $\succ$ is {\em $AC$-compatible} if, for all terms
$t,t',u,u'$ such that $t=_{AC}t'$ and $ u=_{AC}u'$, $t'\succ u'$ iff
$t\succ u$. The relation $\a_{\cR,AC}$ is {\em confluent modulo $AC$}
if ${(\al^*_{\cR,AC}=_{AC}\a^*_{\cR,AC})}\sle
{(\a^*_{\cR,AC}=_{AC}\al^*_{\cR,AC})}$.\vsp[2mm]


\noindent
{\bf Completion modulo AC.} Given a finite set $\cE$ of equations and
an $AC$-compatible reduction ordering $\succ$, completion modulo $AC$
\cite{peterson81jacm} tries to find a finite set $\cR$ of rules such
that:

\begin{lst}{\bu}
\item $\cR$ is included in $\succ$,
\item $\a_{\cR,AC}$ is confluent modulo $AC$,
\item $\cE$ and $\cR\cup\cE_{AC}$ have same theory:
${=_\cE}={=_{\cR\cup\cE_{AC}}}$.
\end{lst}


\begin{dfn}
A theory $\cE$ has a {\em complete presentation} if there is an
AC-com\-patible reduction ordering for which the $AC$-completion of
$\cE_{\neg AC}$ successfully terminates.
\end{dfn}

Many interesting systems have a complete presentation: (commutative)
mo\-no\-ids, (abelian) groups, rings, etc. See
\cite{hullot80thesis,lechenadec86book} for a catalog. Moreover,
there are automated tools implementing completion modulo AC. See for
instance \cite{cime2.02,gaillourdet03cade}.


A term may have distinct $\cR,AC$-normal forms but, by confluence
modulo $AC$, all normal forms are $AC$-equivalent and one can easily
define a notion of normal form for $AC$-equivalent terms
\cite{hullot80thesis}:

\begin{dfn}[$AC$-normal form]
Given an associative and commutative constructor $C$, {\em
$C$-left-combs} (resp. {\em $C$-right-combs}) and their {\em leaves}
are inductively defined as follows:

\begin{lst}{--}
\item If $t$ is not headed by $C$, then $t$ is both a $C$-left-comb
and a $C$-right-comb. The {\em leaves} of $t$ is the one-element list
$\leaves(t)=[t]$.
\item If $t$ is not headed by $C$ and $u$ is a $C$-right-comb,
then $C(t,u)$ is a $C$-right-comb. The {\em leaves} of $C(t,u)$ is the
list $t::\leaves(u)$.
\item If $t$ is not headed by $C$ and $u$ is a $C$-left-comb,
then $C(u,t)$ is a $C$-left-comb. The {\em leaves} of $C(u,t)$ is the
list $\leaves(u)@[t]$, where $@$ is the concatenation.
\end{lst}

\noindent
Let $\orient$ be a function associating a kind of combs (left or
right) to every AC-constructor. Let $\le$ be a total ordering on
terms. Then, a term $t$ is in {\em $AC$-normal form wrt $\orient$ and
$\le$} if:

\begin{lst}{--}
\item Every subterm of $t$ headed by an AC-constructor $C$ is an
$\orient(C)$-comb whose leaves are in increasing order wrt $\le$.
\item For every subterm of $t$ of the form $C(u,v)$ with $C$
commutative but non-associative, we have $u\le v$.
\end{lst}
\end{dfn}


As it is well-known, one can put any term in $AC$-normal form:

\begin{thm}
\label{thm-ac}
Whatever the function $\orient$ and the ordering $\le$ are, every term
$t$ has an $AC$-normal form $t\!\ad_{AC}$ wrt $\orient$ and $\le$, and
$t=_{AC}t\!\ad_{AC}$.
\end{thm}

\begin{prf}
Let $\cA$ be the set of rules obtained by choosing an orientation for
the associativity equations of $\cE_{AC}$ according to $\orient$:

\begin{lst}{--}
\item If $\orient(C)$ is ``left'', then take $C(x,C(y,z))\a C(C(x,y),z)$.
\item If $\orient(C)$ is ``right'', then take $C(C(x,y),z)\a C(x,C(y,z))$.
\end{lst}

$\a_\cA$ is a confluent and terminating relation putting every subterm
headed by an AC-constructor into a comb form according to
$\orient$. Let $\comb$ be a function computing the $\cA$-normal form
of a term. Let now $\sort$ be a function permuting the leaves of combs
and the arguments of commutative but non-associative constructors to
put them in increasing order wrt $\le$. Then, the function
$\sort\circ\comb$ computes the $AC$-normal form of any term and
$\sort(\comb(t))=_{AC} t$.\cqfd\\
\end{prf}


This naturally provides a decision procedure for $AC$-equivalence: the
function $\la xy.\sort(\comb(x))=\sort(\comb(y))$. It follows that
$\cR,AC$-normalization together with $AC$-normalization provides a
valid normalization function, hence the existence of a valid family of
construction functions:

\begin{thm}
\label{thm-ex}
If $\cE$ has a complete presentation, then there exists a valid family
of construction functions.
\end{thm}

\begin{prf}
Assume that $\cE$ has a complete presentation $\cR$. We define the
computation of normal forms as it is generally implemented in
rewriting tools. Let $\step$ be a function making an $\cR,AC$-rewrite
step if there is one, or failing if the term is in normal form. Let
$\norm$ be the function applying $\step$ until a normal form is
reached. Since $\cR$ is a complete presentation of $\cE$, by
definition of the completion procedure, $\sort\circ\comb\circ\norm$ is
a valid normalization function. Thus, by Theorem \ref{thm-g}, the
associated family of construction functions is valid.\cqfd\\
\end{prf}

The construction functions described in the proof are not very
efficient since they are based on rewriting with pattern matching
modulo AC, which is NP-complete \cite{benanav87jsc}, and do not take
advantage of the fact that, by definition of PDTs, they are only
applied to terms already in normal form. We can therefore wonder
whether they can be defined in a more efficient way for some common
equational theories like the ones of Figure \ref{fig-eq}.


\begin{figure*}
\centering\caption{Some common equations on binary constructors\label{fig-eq}}\vsp[2mm]
\begin{tabular}{|c|c|c|c|}\hline
\bf Name & \bf Abbrev & \bf Definition & \bf Example\\\hline
associativity & $Assoc(C)$ & $C(C(x,y),z)=C(x,C(y,z))$ & $(x+y)+z=x+(y+z)$\\\hline
commutativity & $Com(C)$ & $C(x,y)=C(y,x)$ & $x+y=y+x$\\\hline
neutrality & $Neu(C,E)$ & $C(x,E)=x$ & $x+0=x$\\\hline
inverse & $Inv(C,I,E)$ & $C(x,I(x))=E$ & $x+(-x)=0$\\\hline
idempotence & $Idem(C)$ & $C(x,x)=x$ & $x\et x=x$\\\hline
nilpotence & $Nil(C,A)$ & $C(x,x)=A$ & $x\oplus x=\bot$ (exclusive or)\\\hline
\end{tabular}
\end{figure*}


Rewriting provides also a way to check the validity of construction
functions:

\begin{thm}
\label{thm-valid}
If $\cE$ has a complete presentation $\cR$ and $\cF=(f_C)_{C\in\cC}$
is a family such that, for all $C:\s_1..\s_n\pi\in\S$ and terms
$v_i\in Val(\s_i)$, $f_C(v_1..v_n)$ is an $\cR,AC$-normal form of
$C(v_1..v_n)$ in $AC$-normal form, then $\cF$ is valid.
\end{thm}

\begin{prf}
\begin{lst}{--}
\item Correctness. Let $C:\s_1..\s_n\pi\in\S$
and $v_i\in Val(\s_i)$. Since $f_C(v_1..v_n)$ is an $\cR,AC$-normal
form of $C(v_1..v_n)$, we clearly have $f_C(v_1..v_n)=_\cE
C(v_1..v_n)$.

\item Completeness. 
Let $C:\s_1..\s_n\pi\in\S$, $v_i\in Val_\cF(\s_i)$,
$D:\tau_1..\tau_p\pi\in\S$, and $w_i\in Val_\cF(\tau_i)$ such that
$C(v_1..v_n)=_\cE D(w_1..w_p)$. Since $\cR$ is a complete presentation
of $\cE$, $\norm(C(v_1..v_n))=_{AC}\norm(D(w_1..w_p))$. Thus,
$f_C(v_1..v_n)= f_D(w_1..w_p)$.\cqfd\\
\end{lst}
\end{prf}

It follows that rewriting provides a natural way to explain what are
the possible values of an RDT: values are $AC$-normal forms matching
no left hand side of a rule of $\cR$.


\section{Towards efficient construction functions}
\label{sec-genr}


When there is no commutative symbol, construction functions can be
easily implemented by simulating innermost rewriting as follows:

\begin{dfn}[Linearization]
Let $\vpos(t)$ be the set of positions $p\in\pos(t)$ such that $t|_p$
is a variable $x\in\cX$. Let $\r:\vpos(t)\a\cX$ be an injective
mapping and $lin(t)$ be the term obtained by replacing in $t$ every
subterm at position $p\in\vpos(t)$ by $\r(p)$. Let now $Eq(t)$ be the
conjunction of {\tt true} and of the equations $\r(p)=\r(q)$ such that
$t|_p=t|_q$ and $p,q\in\vpos(t)$.
\end{dfn}

\begin{dfn}
\label{dfn-r}
Given a set $\cR$ of rewrite rules, let $\cF(\cR)$ be the family of
construction functions $(f_C)_{C\in\cC}$ defined as follows:

\begin{lst}{\bu}
\item For every rule $l\a r\in\cR$ with $l=C(l_1,\ldots,l_n)$,
add to the definition of $f_C$ the clause {\tt
$lin(l_1),\ldots,lin(l_n)$ when $Eq(l)$ -> $\h{lin(r)}$}, where
$\h{t}$ is the term obtained by replacing in $t$ every occurrence of a
constructor $C$ by a call to its construction function $f_C$.
\item Terminate the definition of $f_C$ by the {\em default clause}
{\tt x -> C(x)}.
\end{lst}
\end{dfn}

\begin{thm}
\label{thm-1}
Assume that $\cE_{AC}=\vide$ and $\cE$ has a complete presentation
$\cR$. Then, $\cF(\cR)$ is valid wrt $\cE$ (whatever the order of the
non-default clauses is).
\end{thm}


We now consider the case of commutative symbols. We are going to
describe a modular way of defining the construction functions by
pursuing our running  example, with the type {\tt exp}. 
Assume that {\tt Plus} is declared to be associative
and commutative only. The construction functions can then be defined
as follows:

{\small\begin{verbatim}
let zero = Zero and one = One and opp x = Opp x

and plus = function
| Plus(x,y), z -> plus (x, plus (y,z))
| x, y -> insert_plus x y

and insert_plus x = function
| Plus(y,_) as u when x <= y -> Plus(x,u)
| Plus(y,t) -> Plus (y, insert_plus x t)
| u when x > u -> Plus(u,x)
| u -> Plus(x,u)
\end{verbatim}}

One can easily see that {\tt plus} does the same job as the function
$\sort\circ\comb$ used in Theorem \ref{thm-ac} but in a slightly more
efficient way since $\cA$-normalization and sorting are interleaved.


Assume moreover that {\tt Zero} is neutral. The AC-completion of
$\{$ {\tt Plus}$(${\tt Zero}$,x)$ $=x\}$ gives $\{$ {\tt Plus}$(${\tt
Zero}$,x)\a x\}$. Hence, if $x$ and $y$ are terms in normal form, then
{\tt Plus}$(x,y)$ can be rewritten modulo AC only if $x=$ {\tt Zero}
or $y= $ {\tt Zero}. Thus, the function {\tt plus} needs to be
extended with two new clauses only:

{\small\begin{verbatim}
and plus = function
| Zero, y -> y
| x, Zero -> x
| Plus(x,y), z -> plus (x, plus (y,z))
| x, y -> insert_plus x y
\end{verbatim}}


Assume now that {\tt Plus} is declared to have {\tt Opp} as
inverse. Then, the completion modulo AC of
$\{$ {\tt Plus}$(${\tt Zero}$,x)=x,$
{\tt Plus}$(${\tt Opp}$(x),x)=$ {\tt Zero}$\}$ 
gives the following well known rules for
abelian groups \cite{hullot80thesis}: 
$\{$ {\tt Plus}$(${\tt Zero}$,x)\a x$,\quad
{\tt Plus}$(${\tt Opp}$(x),x)\a $ {\tt Zero},\quad
{\tt Plus}$(${\tt Plus}$(${\tt Opp}$(x),x),y)\a y$,\quad
{\tt Opp}$(${\tt Zero}$)\a${\tt Zero},\quad
{\tt Opp}$(${\tt Opp}$(x))\a x$,\quad
{\tt Opp}$(${\tt Plus}$(x,y))\a$
{\tt Plus}$(${\tt Opp}$(y),${\tt Opp}$(x))\,\}$.

The rules for {\tt Opp} are easily translated as follows:

{\small\begin{verbatim}
and opp = function
| Zero -> Zero
| Opp(x) -> x
| Plus(x,y) -> plus (opp y, opp x)
| _ -> Opp(x)
\end{verbatim}}

The third rule of abelian groups is called an {\em extension} of the
second one since it is obtained by first adding the context
$Plus([],y)$ on both sides of this second rule,then normalizing the
right hand side. Take now two terms $x$ and $y$ in normal form and
assume that $(x,y)$ matches none of the three clauses previously
defining {\tt plus}, that is, $x$ and $y$ are distinct from {\tt
Zero}, and $x$ is not of the form {\tt Plus}$(x_1,x_2)$. To get the
normal form of {\tt Plus}$(x,y)$, we need to check that $x$ and the
normal form of its opposite {\tt Opp}$(x)$ do not occur in $y$. The
last clause defining {\tt plus} needs therefore to be modified as
follows:

{\small\begin{verbatim}
and plus = function
| Zero, y -> y
| x, Zero -> x
| Plus(x,y), z -> plus (x, plus (y,z))
| x, y -> insert_opp_plus (opp x) y

and insert_opp_plus x y =
  try delete_plus x y
  with Not_found -> insert_plus (opp x) y

and delete_plus x = function
| Plus(y,_) when x < y -> raise Not_found
| Plus(y,t) when x = y -> t
| Plus(y,t) -> Plus (y, delete_plus x t)
| y when y = x -> Zero
| _ -> raise Not_found
\end{verbatim}}

Forgetting about {\tt Zero} and {\tt Opp}, suppose now that {\tt Plus}
is declared associative, commutative and idempotent. The function {\tt
plus} is kept but the {\tt insert} function is modified as follows:

{\small\begin{verbatim}
and insert_plus x = function
| Plus(y,_) as u when x = y -> u
| Plus(y,_) as u when x < y -> Plus(x,u)
| Plus(y,t) -> Plus (y,insert_plus x t)
| u when x > u -> Plus(u,x)
| u when x = u -> u
| u -> Plus(x,u)
\end{verbatim}}

Nilpotence can be dealt with in a similar way.


In conclusion, for various combinations of the equations of Figure
\ref{fig-eq}, we can define in a nice modular way construction
functions that are more efficient than the ones based on rewriting
modulo AC. We summarize this as follows:

\begin{dfn}
A set of equations $\cE$ is a theory of type:
\begin{enumi}{}
\item if $\cE_{AC}=\vide$ and $\cE$ has a complete presentation,
\item if $\cE$ is the union of $\{Assoc(C),Com(C)\}$
with either $\{Neu(C,E),Inv(C,I,E)\}$,
$\{Idem(C)\}$, $\{Neu(C,E),Idem(C)\}$
$\{Nil(C,A)\}$ or $\{Neu(C,E),Nil(C,A)\}$.
\end{enumi}

\noindent
Two theories are disjoint if they share no symbol.
\end{dfn}


Let us give schemes for construction functions for theories of type 2.
A clause is generated only if the conditions {\tt Neu(C,E)}, {\tt
Inv(C,I,E)}, etc. are satisfied. These conditions are not part of the
generated code.

{\small\begin{verbatim}
let f_C = function
| E, x when Neu(C,E) -> x
| x, E when Neu(C,E) -> x
| C(x,y), z when Assoc(C) -> f_C(x,f_C(y,z))
| x, y when Inv(C,I,E) -> insert_inv_C (f_I x) y
| x, y -> insert_C x y

and f_I = function
| E -> E
| I(x) -> x
| C(x,y) -> f_C(f_I y, f_I x)
| x -> I x

and insert_inv_C x y =
  try delete_C x y
  with Not_found -> insert_C (f_I x) y

and delete_C x = function
| Plus(y,_) when x < y -> raise Not_found
| Plus(y,t) when x = y -> t
| Plus(y,t) -> C(y, delete_C x t)
| y when y = x -> E
| _ -> raise Not_found



and insert_C x = function
| C(y,_) as u when x = y & idem -> u
| C(y,t) when x = y & nil -> f_C(A,t)
| C(y,_) as u when x <= y & com -> C(x,u)
| C(y,t) when Com(C) -> C(y, insert_C x t)
| u when x > u & Com(C) -> C(u,x)
| u when x = u & Idem(C) -> u
| u when x = u & Nil(C,A) -> A
| u -> C(x,u)
\end{verbatim}}

\begin{thm}
Let $\cE$ be the union of pairwise disjoint theories of type 1 or
2. Assume that, for all constructor $C$ which theory is of type $k$,
$f_C$ is defined as in Definition \ref{dfn-r} if $k=1$, and as above
if $k=2$. Then, $(f_C)_{C\in\cC}$ is valid wrt $\cE$.
\end{thm}

\begin{prf}
Assume that $\cE=\bigcup_{i=1}^n\cE_i$ where $\cE_1,\ldots,\cE_n$ are
pairwise disjoint theories of type 1 or 2. Whatever the type of
$\cE_i$ is, we saw that $\cE_i$ has a complete presentation
$\cR_i$. Therefore, since $\cE_1,\ldots,\cE_n$ share no symbol, by
definition of completion, the $AC$-completion of $\cE$ successfully
terminates with $\cR=\bigcup_{i=1}^n\cR_i$. Thus, $\a_{\cR,AC}$ is
terminating and $AC$-confluent. Since $\cF=(f_C)_{C\in\cC}$
computes $\cR,AC$-normal forms in $AC$-normal forms, by Theorem
\ref{thm-valid}, $\cF$ is valid.\cqfd\\
\end{prf}

The construction functions of type 2 can be easily extended to deal
with ring or lattice structures (distributivity and absorbance
equations).

More general results can be expected by using or extending results on
the modularity of completeness for the combination of rewrite
systems. The completeness of hierarchical combinations of
non-$AC$-rewrite systems is studied in \cite{rao93fsttcs}. Note
however that the modularity of confluence for $AC$-rewrite systems has
been formally established only recently in \cite{jouannaud06rta}.

Note that the construction function definitions of type 1 or 2 provide
the same results with call-by-value, call-by-name or lazy evaluation
strategy.

The detailed study of the complexity of theses definitions (compared
to AC-rewriting) is left for future work.


\section{The \moca\ system}
\label{sec-moca}

We now describe the \moca\ prototype, a program generator that
implements an extension of \ocaml\ with RDTs. \moca\ parses a special
``.mlm'' file containing the RDT definition and produces a regular
\ocaml\ module (interface and implementation) which provides the
construction functions for the RDT. \moca\ provides a set of keywords
for specifying the equations described in Figure \ref{fig-eq}.

For instance, the RDT {\tt exp} can be defined in \moca\ as follows:

{\small\begin{verbatim}
type exp = private Zero | One | Opp of exp | Plus of exp * exp
  begin associative commutative neutral(Zero) opposite(Opp) end
\end{verbatim}}

\moca\ also features user's arbitrary rules with the construction:
{\small {\tt rule} $pattern$ {\tt ->} $pattern$}. These rules add
extra clauses in the definitions of construction functions generated
by \moca: the LHS $pattern$ is copied verbatim as the pattern of a
clause which returns the RHS $pattern$ considered as an expression
where constructors are replaced by calls to the corresponding
construction functions. Of course, in the presence of such arbitrary
rules, we cannot guarantee the termination or completeness of the
generated code. This construction is thus provided for expert users
that can prove termination and completeness of the corresponding set
of rules. That way, the programmer can describe complex RDTs, even
those which cannot be described with the set of predefined equational
invariants.

\moca\ also accepts polymorphic RDTs and RDTs mutually defined with
record types (but equations between record fields are not yet
available).

The equations of Figure \ref{fig-eq} also support n-ary constructor,
implemented as unary constructors of type {\tt t list -> t}. In this
case, {\tt Plus} gets a single argument of type {\tt exp list}. Normal
forms are modified accordingly and use lists instead of combs. For
instance, associative normal forms get flat lists of arguments: in a
{\tt Plus}$(l)$ expression, no element of $l$ is a {\tt Plus}$(l')$
expression. The corresponding data structure is widely used in
rewriting.

Finally, \moca\ offers an important additional feature: it can
generate construction functions that provide maximally shared
representatives. To fire maximal sharing, just add the {\tt --sharing}
option when compiling the ``.mlm'' file. In this case, the generated
type is slightly modified, since every functional constructor gets an
extra argument to keep the hash code of the term. Maximally shared
representatives have a lot of good properties: not only data size is
minimal and user's memoized functions can be light speed, but
comparison between representatives is turned from a complex recursive
term comparison to a pointer comparison -- a single machine
instruction. \moca\ heavily uses this property for the generation of
construction functions: when dealing with non-linear equations, the
maximal sharing property allows \moca\ to replace term equality by
pointer equality.

\section{Future work}
\label{sec-futur}

We plan to integrate \moca\ to the development environment \focal\
\cite{focal0.3.1}. \focal\ units contain declarations and
definitions of functions, statements and proofs as first-class
citizens. Their compilation produces both a file checkable by the
theorem prover \coq\ \cite{coq8.0} and a \ocaml\ source code. Proofs
are done either within \coq\ or via the automatic theorem prover
\zenon\ \cite{zenon0.4.1}, which issues a \coq\ file when it
successes. Every \focal\ unit has a special field, giving the type of
the data manipulated in this unit. Thus, it would be very interesting
to do a full integration of private/relational data types in \focal,
the proof of correctness of construction functions being done with
\zenon\ or \coq\ and then recorded as a theorem to be used for further
proofs. This should be completed by the integration of a tool on
rewriting and equational theories able to complete equational
presentations, to generate and prove the corresponding lemmas and to
show some termination properties. Some experiments already done within
\focal\ on coupling \cime\ \cite{cime2.02} and \zenon\ give a serious
hope of success.

{\bf Acknowledgments.} The authors thank Claude Kirchner for his
comments on a previous version of the paper.

\bibliographystyle{plain}
\bibliography{biblio} 

\end{document}